\documentclass[a4paper,11pt]{article}
\usepackage[dvips]{graphicx}
\usepackage{amssymb}
\usepackage{amsmath}

\usepackage{cite}

\begin{document}

\title{Singularities of the Casimir Energy for Quantum Field Theories with Lifshitz Dimensions}
\author{V. K. Oikonomou\thanks{
voiko@physics.auth.gr,Vasilis.Oikonomou@mis.mpg.de}\\
Max Planck Institute for Mathematics in the Sciences\\
Inselstrasse 22, 04103 Leipzig, Germany} \maketitle

\begin{abstract}
We study the singularities that the Casimir energy of a scalar field in spacetimes with Lifshitz dimensions exhibits, and provide expressions of the energy in terms of multidimensional zeta functions for the massless case. Using the zeta-regularization method, we found that when the 4-dimensional spacetime has Lifshitz dimensions, then for specific values of the critical exponents, the Casimir energy is singular, in contrast to the non-Lifshitz case. Particularly we found that when the value of the critical exponent is $z=2$, the Casimir energy is singular, while for $z\geq 3$ the Casimir energy is regular. In addition, when flat extra dimensions are considered, the critical exponents of the Lifshitz dimensions affect drastically the Casimir energy, introducing singularities that are absent in the non-Lifshitz case. We also discuss the Casimir energy in the context of braneworld models and the perspective of Lifshitz dimensions in such framework.
\end{abstract}

\section*{Introduction}

During the last decades, the Casimir effect has played a prominent role in various research areas of theoretical physics (see for example \cite{cas1,cas2,cas3,cas4,cas6,cas8,cas9,cas10} and references therein). The first theoretical study was done by Casimir \cite{cas1}, in 1948, who predicted an attractive force between two neutral perfectly conducting parallel plates. The Casimir effect is a manifestation of the vacuum corresponding to a quantum field theory. Owing to the vacuum fluctuations of the
electromagnetic quantum field, the parallel plates attract each other. Moreover, the boundaries alter the quantum field boundary conditions and as a result the plates interact. The geometry of the boundaries have a strong effect
on the Casimir energy and Casimir force. The calculations of the Casimir energy and the corresponding force were generalized to include other
quantum fields such as fermions, bosons and scalar fields
making the Casimir effect study an important
ingredient of many theoretical physics subjects such as string
theory, cosmology e.t.c.
The technological applications of the Casimir force are of invaluable importance, for
instance in nanotubes, nano-devices and generally in
microelectronic engineering \cite{micro}. Indeed an attractive or repulsive Casimir force can
lead to the instability or even destruction of such a
micro-device. Hence, studying various geometrical and
material configurations will enable us to have control over the
Casimir force.

\noindent The Casimir effect was verified experimentally \cite{exper,exper1} rendering the physics of such studies very valuable due to the theoretical outcomes of these studies. The Casimir energy studies have been done for curved spacetimes, for topologically non trivial backgrounds and for various geometrical configurations and physical setups \cite{cas1,cas2,cas3,cas4,cas6,cas8,cas9,cas10}.  The applications of such calculations are numerous, constraining even cosmological models. Moreover, various models are put in test, for example the size and the
shape of extra dimensions are constrained from Casimir energy calculations \cite{extracas,extracas1}. In view of these applications, every consistent quantum field theory is severely constrained by Casimir energy and Casimir force measurements.

\noindent Lifshitz type quantum field theories \cite{Lif,Lif1,Lif2,Lif3,Lif4,Lif5,Lif6,Lif7,Lif8,Lif9,Lif10,Lif11,Lif12,Lif13,Lif14,Lif15,Lif16,Lif17,Lif18,Lif19,Lif20,Lif21,Lif22} serve as Lorentz violating field theories with remarkable properties. These theories have their origin in condensed matter physics \cite{cond} where a Lifshitz critical point is defined as a point in a phase diagram where three phases of a condensed matter system meet. Condensed matter systems exhibiting a Lifshitz point have an intrinsic space anisotropy, which is quantified in terms of the existence of two different correlation lengths in the anisotropic space dimensions. Lifshitz quantum field theories have been studied in flat and curved background and, although being Lorentz violating theories, the renormalization of the various loop integrals is improved, with the last being the most appealing attribute of these gauge theories. In addition, the class of the renormalizable interactions is sufficiently enriched, since the
Lifshitz type
operators that appear in a Lagrangian are higher derivatives of the quantum
fields with mass dimensions. Another appealing feature is that, within the Lifshitz theoretical framework, dynamical mass generation naturally occurs, as a consequence of the dimensionfull couplings of Lifshitz operators. We shall study a  massless scalar field, in the context of Lifshitz type dimensions, but in two different situations.

\noindent Since the Casimir energy has proved to be a very important property of any quantum field theory, we shall compute the Casimir energy of a scalar field, in four dimensional spacetime with Lifshitz anisotropic extra dimensions and also the Casimir energy in four dimensions with anisotropic scaling between space and time. We extend the last framework in the case which each space dimension is anisotropic. What we are mainly interested in, is the singularity structure of the Casimir energy in all the aforementioned cases. We want to explore if the introduction of a Lifshitz anisotropy in space introduces any new singularities. Since the Casimir energy is a benchmark for any viable quantum field theory, such a study is very important for the intrinsic validity of any quantum field theory.

\noindent In this paper we shall calculate the Casimir energy for a massless scalar field confined in a three dimensional box with Dirichlet boundaries. It will be assumed that some of the dimensions are Lifshitz in various setups. As we shall demonstrate, the quantum theory with critical exponent $z=2$ yields a singular Casimir energy even though we use the zeta-regularization method. This is in contrast to the non-Lifshitz case and also for Lifshitz quantum theories with $z\geq 3$, where the Casimir energy is finite. Hence, theories with $z=2$ are put into question. In addition we shall assume that the 4-dimensional spacetime has Lifshitz extra dimensions. In this framework the four dimensional Minkowski spacetime Lorentz invariance is protected, while each extra dimension is anisotropic. However this anisotropy has a radical impact on the field theory Lagrangian, since it allows a quite large number of higher order derivatives of the fields. Nevertheless, when the theory is assumed to ''sit'' on a Lifshitz point, the only derivative terms that survive are the higher order ones, since the lower order derivatives at the Lifshitz point become irrelevant. This construction is very similar to the original condensed matter
construction and additionally it doesn't affect the four dimensional Lorentz invariance. As we shall see, extra dimensions bring about singular terms that are absent in the non-Lifshitz case. Finally, since the Kaluza-Klein theories with flat extra dimensions are not considered to be realistic, we shall comment on the calculation of the 	Casimir energy in Randall-Sundrum like spacetimes and more generally, in the context of braneworld scenario. We shall assume the spacetime is curved and that gravity is coupled conformally to the scalar field. Particularly, the spacetime will shall use is a generalization of AdS spacetime with Lifshitz scaling invariance. As we shall evince, the calculation of the scalar Casimir energy in such backgrounds is not an easy task, since the Einstein equations have no solutions of Lifshitz type.

\noindent This letter is organized as follows:  In section 1 we compute the Casimir energy for 4-dimensional Minkowski spacetime with anisotropic scaling between space and time. The massless scalar field is considered to be confined in a box, and also satisfies Dirichlet boundary conditions on the boundaries. We also consider the case where the three space dimensions are anisotropic. In section 2 we calculate the Casimir energy of a massless scalar field in a Minkowski spacetime background with a Lifshitz extra dimension and investigate how the singularities of the Casimir energy are affected by the anisotropic scaling of the extra dimension. Moreover we present the singularities of the massless scalar field Casimir energy, in Minkowski spacetime with two anisotropic extra dimensions. We discuss on the calculation of the Casimir energy in the context of braneworld scenario in section 3. The conclusions with a discussion on the results follow in the end of this letter.

\section{Casimir Energy for Scalar Field with Three Spatial Lifshitz Dimensions}

\subsection*{Massless Scalar Case-Identical Scaling of the Space Dimensions}

We shall start with the study of the Casimir energy for a scalar field in three dimensional spacetime but with the space and time scaled differently. Particularly, we assume that the mass dimension of the spacetime coordinates is:
\begin{equation}\label{threecase}
[t]=-z,{\,}{\,}{\,}{\,}{\,}[x_i]=-1
\end{equation}
with $i=1,2,3$. This case has been studied thoroughly and appears quite frequently in the literature, see for example reference \cite{Lif8}. Relation (\ref{threecase}) stems from the Lifshitz scaling of the time coordinate in reference to that of the three spatial coordinates, namely:
\begin{equation}\label{scalethree}
x_{i}\rightarrow b x_{i},{\,}{\,}{\,}{\,}{\,}t\rightarrow b^{z}t
\end{equation}
As we mentioned in the introduction, although such a scaling is different in spirit regarding the original idea of Lifshitz \cite{cond}, the renormalization properties of such scalar theories are remarkable rendering such theories really valuable, although Lorentz violating. We want to compute the Casimir energy in the case the system is confined in a box, with lengths, $0\leq x_i\leq L_i$ and $i=1,2,3$.
Assuming relation (\ref{threecase}), the action of the massless scalar field reads,
\begin{align}\label{action2}
&\mathcal{S}=\int d\mathrm{t}\int_0^{L_1}\int_0^{L_2}\int_0^{L_3}d\mathrm{x_1}d\mathrm{x_2}d\mathrm{x_3}
\\ \notag & \times \Phi^*(t,x_i)\Big{(}(\partial^{t})^2-(-\partial_{x_1}^2)^{z}-(-\partial_{x_2}^2)^{z}-(-\partial_{x_3}^2)^{z}\Big{)}\Phi (t,x_i)
\end{align}
The scalar field is considered to obey Dirichlet boundary conditions at the boundary of the box,
\begin{align}\label{periodicitydir}
& \Phi(x_1,x_2,0)=\Phi(x_1,x_2,L_3)=0
\\ \notag &\Phi(x_1,0,x_3)=\Phi(x_1,L_2,x_3)=0
\\ \notag &\Phi(0,x_2,x_3)=\Phi(L_1,x_2,x_3)=0
\end{align}
and as a consequence the Casimir energy is equal to:
\begin{equation}\label{casimiren105}
\mathcal{E}_{z_1,z_2,z_3}=
\sum_{n_1,n_2,n_3=1}^{\infty}\Big{[}\Big{(}\frac{2n_1\pi}{L_1}\Big{)}^{2z}+\Big{(}\frac{2n_2\pi}{L_2}\Big{)}^{2z}+\Big{(}\frac{2n_3\pi}{L_3}\Big{)}^{2z}\Big{]}^{-s}
\end{equation}
The expression for the Casimir energy (\ref{casimiren105}), can be written in terms of a multidimensional zeta function \cite{elizalde}, namely in terms of the function,
\begin{align}\label{3dimensionalMpre}
&\mathcal{M}_3(s_1;a_1,a_2,a_3;m_1,m_2,m_3)=\sum_{n_1,n_2,n_3=1}^{\infty}(a_1n_1^{m_1}+a_2n_2^{m_2}+a_3n_3^{m_3})^{-s_1}
\\ \notag & \simeq \frac{1}{a_3^{s_1}\Gamma(s_1)}\Big{(}\sum_{k_1,k_2=0}^{\infty}\frac{(-b_1)^{k_1}}{k_1!}\frac{(-b_2)^{k_2}}{k_2!}
\\ \notag & \times \Gamma(s_1+k_1+k_2)\zeta(-m_1k_1)\zeta(-m_2k_2)\zeta(m_3(s_1+k_1+k_2))\Big{)}
\\ \notag & + \frac{\Gamma (1/m_2)}{m_2b_2^{1/m_2}}\sum_{k_1=0}^{\infty}\frac{(-b_1)^{k_1}}{k_1!} \Gamma(s_1+k_1-1/m_2)\zeta(-m_1k_1)\zeta(m_3(s_1+k_1-1/m_2))
\\ \notag & + \frac{\Gamma (1/m_1)}{m_1b_1^{1/m_1}}\sum_{k_2=0}^{\infty}\frac{(-b_2)^{k_2}}{k_2!} \Gamma(s_1+k_2-1/m_1)\zeta(-m_2k_2)\zeta(m_3(s_1+k_2-1/m_1))
\\ \notag & +\frac{\Gamma (1/m_1)\Gamma (1/m_2)}{m_1b_1^{1/m_1}m_2b_2^{1/m_2}}\Gamma(s_1-1/m_1-1/m_2)\zeta(m_3(s_1-1/m_1-1/m_2))
\end{align}
with $b_j=a_j/a_3$, $j=1,2$. Thereby, the Casimir energy equals to:
\begin{equation}\label{casdir1}
\mathcal{E}_{z_1,z_2,z_3}=\mathcal{M}_3(s_1;a_1,a_2,a_3;m_1,m_2,m_3)
\end{equation}
with the obvious identification,
\begin{align}\label{z1z2dir}
& a_1=\Big{(}\frac{\pi}{L_1}\Big{)}^{2z}, {\,}{\,}{\,}a_2=\Big{(}\frac{\pi}{L_2}\Big{)}^{2z}, {\,}{\,}{\,}a_3=\Big{(}\frac{\pi}{L_3}\Big{)}^{2z}\\
\notag & m_1=m_2=m_3=2z, {\,}{\,}{\,}s_1=-\frac{1}{2},
\end{align}
Before starting examining the singularities of the Lifshitz Casimir energy, let us point that the non-Lifshitz case ($z=1$ case) contains no singularities. 

\noindent When the critical exponent takes values $z\geq 3$ the Casimir energy is finite, as can be easily checked. In that case, all the terms of (\ref{3dimensionalMpre}) are free of singularities. Hence within the theoretical framework of Lifshitz dimensions, it is possible to obtain a finite Casimir energy. Furthermore, since $\zeta(-2zk_i)=0$ for every $z\geq 3$ and $k_1,k_2\geq 1$, the Casimir energy takes a very simple form, namely:
\begin{align}\label{3dimensionalMpresimple}
&\mathcal{E}_{z_1,z_2,z_3}\simeq \frac{1}{a_3^{s_1}\Gamma(s_1)}
\frac{\Gamma(-1/2)}{480}
\\ \notag & - \frac{\Gamma (1/m_2)}{m_2b_2^{1/m_2}} \frac{\Gamma(s_1-1/m_2)}{2}\zeta(m_3(s_1-1/m_2))
\\ \notag & - \frac{\Gamma (1/m_1)}{m_1b_1^{1/m_1}} \frac{\Gamma(s_1-1/m_1)}{2}\zeta(m_3(s_1-1/m_1))
\\ \notag & +\frac{\Gamma (1/m_1)\Gamma (1/m_2)}{m_1b_1^{1/m_1}m_2b_2^{1/m_2}}\Gamma(s_1-1/m_1-1/m_2)\zeta(m_3(s_1-1/m_1-1/m_2))
\end{align}
where the parameters $m_i$ and $s_1$ are given by relation (\ref{z1z2dir}). The finiteness of the Casimir energy without any other renormalization procedure is of particular importance. This is due to the fact that the Casimir energy is a basic feature of any quantum theory, and actually is a manifestation of the quantum vacuum. Hence, the Lifshitz type theories with identical scaling between the space dimensions, but different between space and time, yield a finite Casimir energy for $z\geq 3$. This result probably stems from the fact that since the renormalization properties of Lifshitz type theories are refined, this fact is materialized in the Casimir energy too, at least for $z\geq 3$.

\noindent However there is a value of the critical exponent for which the Casimir energy contains singularities, even though we use the zeta-regularization method. The problematic case for the Lifshitz type theories under study, is realized for $z=2$, or equivalently for $m_1=m_2=m_3=4$. Let us present the singularities of the Casimir energy in this case. In equation (\ref{3dimensionalMpre}) for $z=2$, the fourth and fifth line terms are singular. In addition, the last line term is singular for $z=2$. Hence, the $z=2$ case results corresponds to a singular theory, in comparison to the finite non-Lifshitz case ($z=1$). Such a theory would require new renormalization terms in order to extract the finite Casimir energy out of the infinite expression. We refrain from going into further details since our aim was just to report on the fact that the Casimir energy becomes finite for some values of the critical exponents and in addition to examine if the Casimir energy in the Lifshitz theory context, contains singular terms. As we saw, this happens only for $z=2$. This result is generally strange, since someone expects that since the renormalization properties of Lifshitz theories are refined, the total Casimir energy should be finite for all critical exponents. As we saw, this is not true for $z=2$, thus putting in question this particular case.

\subsection*{Massless Scalar Case-non Identical Scaling of the Space Dimensions}

We now address the problem in which the four dimensional spacetime has three spatial dimensions that are Lifshitz and one temporal dimension which is non-Lifshitz. The dimensions are scaled in the following way:
\begin{equation}\label{scalethreeonly}
x_{i}\rightarrow b^{z_i} x_{i},{\,}{\,}{\,}{\,}{\,}t\rightarrow b t
\end{equation}
with $i=1,2,3$. We assumed that the critical exponents of the spatial dimensions scalings are different. Therefore, the mass dimensions of the spacetime dimensions are:
\begin{equation}\label{newscaledimnon}
[t]=-1,{\,}{\,}{\,}[x_1]=-z_1,{\,}{\,}{\,}[x_2]=-z_2,{\,}{\,}{\,}[x_3]=-z_3
\end{equation}
The corresponding action of the free massless scalar field in this spacetime is,
\begin{align}\label{action2}
&\mathcal{S}=\int d\mathrm{t}\int_0^{L_1}\int_0^{L_2}\int_0^{L_3}d\mathrm{x_1}d\mathrm{x_2}d\mathrm{x_3}
\\ \notag & \times \Phi^*(t,x_i)\Big{(}(\partial^{t})^2-q_1^{2(z_1-1)}(-\partial_{x_2}^2)^{z_2}-q_2^{2(z_2-1)}(-\partial_{x_2}^2)^{z_2}-q_3^{2(z_3-1)}(-\partial_{x_3}^2)^{z_3}\Big{)}\Phi (t,x_i)
\end{align}
with $i=1,2,3$. We confine the scalar field in a box, like in the previous section, and we impose Dirichlet boundary conditions at the boundaries of the three dimensional box. These are,
\begin{align}\label{periodicity}
& \Phi(x_1,x_2,0)=\Phi(x_1,x_2,L_3)=0
\\ \notag &\Phi(x_1,0,x_3)=\Phi(x_1,L_2,x_3)=0
\\ \notag &\Phi(0,x_2,x_3)=\Phi(L_1,x_2,x_3)=0
\end{align}
Accordingly, the Casimir energy reads,
\begin{equation}\label{casimiren105}
\mathcal{E}_{z_1,z_2,z_3}^A=
\sum_{n_1,n_2,n_3=1}^{\infty}\Big{[}q_1^{2(z_1-1)}\Big{(}\frac{2n_1\pi}{L_1}\Big{)}^{2z_1}+q_2^{2(z_2-1)}\Big{(}\frac{2n_2\pi}{L_2}\Big{)}^{2z_2}+q_3^{2(z_3-1)}\Big{(}\frac{2n_3\pi}{L_3}\Big{)}^{2z_3}\Big{]}^{-s}
\end{equation}
and using the multidimensional zeta function (\ref{3dimensionalMpre}), it can be written as follows,
\begin{equation}\label{casdir1}
\mathcal{E}_{z_1,z_2,z_3}^A=\mathcal{M}_3(s_1;a_1,a_2,a_3;m_1,m_2,m_3)
\end{equation}
with the parameters being equal to,
\begin{align}\label{z1z2newalign}
& a_1=q_1^{2(z_1-1)}\Big{(}\frac{\pi}{L_1}\Big{)}^{2z_1}, {\,}{\,}{\,}a_2=q_2^{2(z_2-1)}\Big{(}\frac{\pi}{L_2}\Big{)}^{2z_2}, {\,}{\,}{\,}a_3=q_3^{2(z_3-1)}\Big{(}\frac{\pi}{L_3}\Big{)}^{2z_3}\\
\notag & m_1=2z_1, {\,}{\,}{\,}m_2=2z_2, {\,}{\,}{\,}m_3=2z_3, {\,}{\,}{\,}
s_1=-\frac{1}{2}
\end{align}
Let us present the singularities in the present case, by investigating equation (\ref{3dimensionalMpre}). To start with, consider the term in the third line, which is always regular (recall $s_1=-1/2$). We focus our interest to the terms appearing in the fourth, fifth and sixth line of equation  (\ref{3dimensionalMpre}). For $z_1=z_2=1$ and $z_3\geq 2$, the fourth and fifth term are singular (gamma function poles) while the sixth is regular. In addition, for $z_3\geq 3$ and $z_1\geq 2$ and $z_2=1$, the fourth term is singular, the fifth term is regular and the sixth is regular. Consider now the case with $z_3\geq 1$ and $z_1=z_2=2$. The fourth term is regular, the fifth term is regular and the sixth is singular.

\noindent But as in the previous section, there exist some values of the critical exponents, for which the Casimir energy is completely finite. Indeed, when $z_1,z_2>2$ and $z_3\geq 1$, all the terms of  (\ref{3dimensionalMpre}) are regular. So when the total spacetime dimension is four, Lifshitz theories result to a completely finite Casimir energy for a wide range of values that the critical exponents take. This result further supports the physical appeal of Lifshitz theories, at least when renormalization issues are considered. In the next section we focus our interest on spacetimes with extra dimensions.

\section{Casimir Energy of Massless Scalar Field for Spacetimes with Lifshitz Extra Dimensions}

\subsection{Casimir Energy for Massless Scalar Field in $M_4\times S_{z_1}^1$}

In this section we compute the Casimir energy for a scalar quantum field in $M_4\times S_{z_1}^1$ spacetime, with $M_4$ denoting the four dimensional Minkowski spacetime, while $S_{z_1}^1$ is the extra dimension with anisotropic scaling. This anisotropy is quantified in terms of the dynamical critical exponent $z_1$, and this means that the scaling of the dimensions is of the form:
\begin{equation}\label{scale}
x_{\mu}\rightarrow b x_{\mu},{\,}{\,}{\,}{\,}{\,}y_1\rightarrow b^{z_1}y_1
\end{equation}
In the above equation, $\mu=0,1,2,3$ and $x_{\mu}$ denotes the Minkowski spacetime coordinates, while $y_1$ denotes the coordinate describing the extra dimension, which is a circle. The action of the free massless scalar field in $M_4\times S_{z_1}^1$ is equal to,
\begin{align}\label{action1}
\mathcal{S}=\int d\mathrm{t}d^{D-1}p\int_0^{2\pi R_1}d\mathrm{y_1}\Phi^*(x_{\mu},y_1)\Big{(}\partial^{\mu}\partial_{\mu}-q_1^{2(z_1-1)}(-\partial_{y_1}^2)^{z_1}\Big{)}\Phi (x_{\mu},y_1)
\end{align}
Note that in the above action we defined the quantum field theory on the Lifshitz critical point $z_1$, thus only the highest order derivatives appear. The mass dimensions of the coordinates are equal to:
\begin{equation}\label{scale1}
[x_{\mu}]=-1,{\,}{\,}{\,}{\,}{\,}[y_1]= -z_1
\end{equation}
and this explains the necessity of the dimensionfull mass parameter $q$. We assume that the scalar field satisfies periodic boundary conditions in the Lifshitz extra dimension, $\Phi(x_{\mu},0)=\Phi(x_{\mu},2\pi R_1)$, with $R_1$ the radius of the extra dimension.
We shall use the dimensional regularization technique in order to calculate the Casimir energy. Therefore, we suppose that $D$ is the total ordinary space dimensionality, hence the Casimir energy reads,
\begin{equation}\label{casimiren1}
\mathcal{E}=\frac{1}{(2\pi)^{D-1}}\int
\mathrm{d}^{D-1}p\sum_{n_1=-\infty}^{\infty}\Big{[}\sum_{k=1}^{D-1}p_k^2+q_1^{2(z_1-1)}\Big{(}\frac{n_1\pi}{R_1}\Big{)}^{2z_1}\Big{]}^{-s}
\end{equation}
It is crucial in the end to replace $s=-1/2$ and $D=4$. Notice that the integration is performed for the non-compact space dimensions, which are $D-1$.
In order to distinguish between contributions to the energy, coming from the extra dimensions and contributions coming from the $D-1$ dimensions, we rewrite the Casimir energy in the form:
\begin{equation}\label{1dcane}
\mathcal{E}=\mathcal{E}^{D-1}+\mathcal{E}_{z_1}
\end{equation}
The term $\mathcal{E}^{D-1}$ is the $D-1$-dimensional contribution to the Casimir energy, and $\mathcal{E}_{z_1}$ denotes the extra dimensional contribution to the Casimir energy. These two are equal to:
\begin{align}\label{newcas}
& \mathcal{E}^{D-1}=\frac{1}{(2\pi)^{D-1}}\int
\mathrm{d}^{D-1}p\Big{[}\sum_{k=1}^{D-1}p_k^2\Big{]}^{-s} \\ \notag &
\mathcal{E}_{z_1}=\frac{1}{(2\pi)^{D-1}}\int
\mathrm{d}^{D-1}p\sum_{n_1=-\infty}^{\infty '}\Big{[}\sum_{k=1}^{D-1}p_k^2+q_1^{2(z_1-1)}\Big{(}\frac{n_1\pi}{R_1}\Big{)}^{2z_1}\Big{]}^{-s}
\end{align}
where the prime in the summation indicates the omission from the sum of the $n_1=0$ term. We shall be mostly interested in the contribution coming from the extra dimension, and particularly we shall present the singularities of the Casimir energy and the effect of the critical exponent $z_1$ on these singularities, in comparison to the non-Lifshitz case ($z_1=1$). Upon integrating over the continuous dimensions and using the zeta function regularization method \cite{cas1,cas2,cas3,cas4,cas6,cas8,cas9,cas10}, the extra dimensional contribution to the energy is, equal to:
\begin{align}\label{pord12}
&\mathcal{E}_{z_1}=\frac{2}{(2\pi)^{D-1}}\pi^{\frac{D-1}{2}}\frac{\Gamma(s-\frac{D-1}{2})}{\Gamma(s)}q_1^{-2(z_1-1)(s-\frac{D-1}{2})}\Big{(}\frac{\pi}{R_1}\Big{)}^{-2z_1(s-\frac{D-1}{2})}\zeta
\Big{(}2z_1(s-\frac{D-1}{2})\Big{)}
\end{align}
Obviously, the zeta function is singular when its
argument, namely $2z_1(s-\frac{D-1}{2})$, is equal to one, which cannot be true for any value of the critical exponent $z_1$. Recall that we take
$D=4$ and $s=-1/2$ in order to recover the Minkowski space with an extra dimension case. We focus on the extra dimensions
dependent terms of the Casimir energy, as we already mentioned. The case $z_1=1$ corresponds to the non-Lifshitz field
theory case, in which case, $\mathcal{E}_{z_1}$ is finite. It is
easy to verify that any value of $z_1$, with $z_1\geq 2$ renders
$\mathcal{E}_{z_1}$ infinite, with the singularity being a pole of
the gamma function, that is:
\begin{equation}\label{sing1}
\mathcal{E}_{z_1}\sim \Gamma (-2)\zeta (-4z_1)
\end{equation}
Therefore, we conclude that the Casimir energy is infinite and new
regularization terms are probably needed to render the Casimir energy finite, a fact that does
not happen for the non-Lifshitz field theories. Hence the scalar field theory defined on the critical point $z_1$, in the spacetime $M_4\times S_{z_1}^1$, with $z_1\geq 2$ introduces singularities that do not exist in the non-Lifshitz case. A final notice regarding the singularities. For the case $z_1=1$, the product of the gamma function with the zeta function satisfy the reflection formula
\begin{equation}\label{reflection}
\Gamma(\frac{z}{2})\zeta (z)=\pi^{z-1/2}\Gamma
(\frac{1-z}{2})\zeta (1-z)
\end{equation}
which renders the extra dimensions Casimir energy contribution finite. Note that for $z_1\geq 2$, the identity above does not hold, so the expression for the Casimir energy is singular.

\noindent The existence of a larger number of Lifshitz extra dimensions further introduces new singularities in the Casimir energy. This is the result of the next section.

\subsection{Casimir Energy for Massless Scalar Field $M_4\times S_{z_1}^1\times S_{z_2}^1$}

We now calculate the Casimir energy for a spacetime of the form, $M_4\times S_{z_1}^1\times S_{z_2}^1$, with the extra dimensions being Lifshitz type dimensions. The dimensions have scalings that this time are of the form,
\begin{equation}\label{scalenew}
x_{\mu}\rightarrow b x_{\mu},{\,}{\,}{\,}{\,}{\,}y_1\rightarrow b^{z_1}y_1,{\,}{\,}{\,}{\,}{\,}y_2\rightarrow b^{z_2}y_2
\end{equation}
The action of the free massless scalar field with two periodic Lifshitz extra dimensions is equal to,
\begin{align}\label{action2}
&\mathcal{S}=\int d\mathrm{t}d^{D-1}p\int_0^{2\pi R_1}\int_0^{2\pi R_2}d\mathrm{y_1}d\mathrm{y_2}
\\ \notag & \times \Phi^*(x_{\mu},y_i)\Big{(}\partial^{\mu}\partial_{\mu}-q_1^{2(z_1-1)}(-\partial_{y_1}^2)^{z_1}-q_2^{2(z_2-1)}(-\partial_{y_2}^2)^{z_2}\Big{)}\Phi (x_{\mu},y_i)
\end{align}
with $i=1,2$, $\mu =0,1,2,3$ and $q_1,q_2$ mass parameters introduced in order the dimensionality of the derivative terms are correct. In this case, the mass dimensions of the coordinates is:
\begin{equation}\label{scale1}
[x_{\mu}]=-1,{\,}{\,}{\,}{\,}{\,}[y_i]= -z_i
\end{equation}
We deploy periodic boundary conditions for the scalar field in the extra dimensions:
\begin{equation}\label{periodicity}
\Phi(x_{\mu},0,y_2)=\Phi(x_{\mu},2\pi R_1,y_2),{\,}{\,} \Phi(x_{\mu},y_1,0)=\Phi(x_{\mu},y_1,2\pi R_2)
\end{equation}
with $R_1,R_2$ denoting the radius of each extra dimension. Using the dimensional regularization technique, the Casimir energy reads,
\begin{equation}\label{casimiren105}
\mathcal{E}=\frac{1}{(2\pi)^{D-1}}\int
\mathrm{d}^{D-1}p\sum_{n_1,n_2=-\infty}^{\infty}\Big{[}\sum_{k=1}^{D-1}p_k^2+q_1^{2(z_1-1)}\Big{(}\frac{n_1\pi}{R_1}\Big{)}^{2z_1}+q_2^{2(z_2-1)}\Big{(}\frac{n_2\pi}{R_2}\Big{)}^{2z_2}\Big{]}^{-s}
\end{equation}
where in the end $s=-1/2$ and $D=4$ as previously. The Casimir energy can be written in the form,
\begin{equation}\label{1dcane2}
\mathcal{E}=\mathcal{E}^{D-1}+\mathcal{E}_{z_1,z_2}+\mathcal{E}_{z_1,z_2}(n_1)+\mathcal{E}_{z_1,z_2}(n_2)
\end{equation}
with $\mathcal{E}_{z_1,z_2}$, $\mathcal{E}_{z_1,z_2}(n_1)$ and $\mathcal{E}_{z_1,z_2}(n_2)$ the extra dimensional contributions to the Casimir energy, which will be the subject of our study in this section. Using the multidimensional zeta function \cite{elizalde} the term $\mathcal{E}_{z_1,z_2}$, is equal to,
\begin{align}\label{p2008ordoulis12}
&\mathcal{E}_{z_1,z_2}=\frac{1}{(2\pi)^{D-1}}\pi^{\frac{D-1}{2}}\frac{1}{2\Gamma(s)}\Big{(}
\sum_{k=0}^{\infty }\frac{\Gamma
(s_1+k)}{k!}\frac{(-a_1)^k}{a_2^{s_1+k}}\zeta(-m_1k)\zeta
(m_2(s_1+k)) \\ \notag & +\frac{\Gamma(1/m_1)\Gamma(s_1-1/m_1)}
{m_1m_2^{s_1}}\Big{(}\frac{a_2}{a_1}\Big{)}^{1/m_1}\zeta
(m_2(s_1-1/m_1))\Big{)}
\end{align}
where in the present case the parameters are equal to:
\begin{align}\label{z1z2}
& a_1=q_1^{2(z_1-1)}\Big{(}\frac{\pi}{R_1}\Big{)}^{2z_1}, {\,}{\,}{\,}a_2=q_2^{2(z_2-1)}\Big{(}\frac{\pi}{R_2}\Big{)}^{2z_2}\\
\notag & m_1=2z_1, {\,}{\,}{\,}m_2=2z_2, {\,}{\,}{\,}
s_1=s-\frac{D-1}{2}
\end{align}
The other two terms, $\mathcal{E}_{z_1,z_2}(n_1)$ and
$\mathcal{E}_{z_1,z_2}(n_2)$ can be written in terms of the
Riemann zeta function, as we saw in the previous section,
\begin{align}\label{zazb}
&\mathcal{E}_{z_1,z_2}(n_1)=\frac{2}{(2\pi)^{D-1}}\pi^{\frac{D-1}{2}}\frac{\Gamma(s-\frac{D-1}{2})}{\Gamma(s)}q_1^{-2(z_1-1)(s-\frac{D-1}{2})}\Big{(}\frac{\pi}{R_1}\Big{)}^{-2z_1(s-\frac{D-1}{2})}\zeta
\Big{(}2z_1(s-\frac{D-1}{2})\Big{)}\\ \notag &
\mathcal{E}_{z_1,z_2}(n_2)\frac{2}{(2\pi)^{D-1}}\pi^{\frac{D-1}{2}}\frac{\Gamma(s-\frac{D-1}{2})}{\Gamma(s)}q_2^{-2(z_2-1)(s-\frac{D-1}{2})}\Big{(}\frac{\pi}{R_2}\Big{)}^{-2z_2(s-\frac{D-1}{2})}\zeta
\Big{(}2z_2(s-\frac{D-1}{2})\Big{)}
\end{align}

\subsection*{Singularities for Various $z_1,z_2$ Values}

Let us present now the singularities that the extra dimensions related terms exhibit. The singularities of the terms $\mathcal{E}_{z_1,z_2}(n_1)$ and
$\mathcal{E}_{z_1,z_2}(n_1)$ are similar to the singularities we
analyzed in the previous section, so for $z_1\geq 2$ and $z_2\geq 2$ respectively, the aforementioned terms are singular. Particularly, we found that when
$z_1\geq 2$ and $z_2\geq 2$, the $\mathcal{E}_{z_1,z_2}(n_1)$ and
$\mathcal{E}_{z_1,z_2}(n_2)$ terms contain singularities which are
of the form:
\begin{equation}\label{sing1mass2}
\mathcal{E}_{z_1,z_2}(n_1)\sim \Gamma
(-2),{\,}{\,}{\,}\mathcal{E}_{z_1,z_2}(n_2)\sim \Gamma (-2)
\end{equation}
The term $\mathcal{E}_{z_1,z_2}$ contains various singularities
for specific values of the summation parameter $k$, which depend on the Lifshitz parameters $z_1$ and
$z_2$. Remarkably, the singularity depends only on the value of $z_2$. The singularities are of the following type, and for the
corresponding $k$ values:
\begin{align}\label{zazb1}
& k=0,{\,}{\,}{\,}\Gamma (-2)\\ \notag &
k=1,{\,}{\,}{\,}\Gamma (-1) \\ \notag & k=2,{\,}{\,}{\,}\Gamma (0)
\end{align}
Therefore, the Lifshitz theory Casimir energy, contains additional
singularities which are the singularities of the functions
$\mathcal{E}_{z_1,z_2}(n_1)$, $\mathcal{E}_{z_1,z_2}$ and $\mathcal{E}_{z_1,z_2}(n_2)$ for
$z_1,z_2\geq 2$. In turn, these singularities would need special
treatment in order the resulting expression of the energy is finite, thus putting in
question such extra dimensional spaces. Note that when $z_1,z_2\geq 2$ there are five sources of singularities and also that when $z_2=1$, and $z_1\geq 2$, the only new singularity is contained in the term $\mathcal{E}_{z_1,z_2}(n_1)$, since all other terms do not contain singularities, exactly as in the non-Lifshitz case.

\section{A Discussion on Lifshitz Extra Dimensions in Randall-Sundrum Spacetimes and Scalar Casimir Energy}

In the previous section we calculated the Casimir energy in the case the total space time is the topological product of four dimensional Minkowski spacetime times circular flat extra dimensions. However, models with flat compact extra dimensions and more generally Kaluza-Klein theories with flat extra dimensions, are not considered to be so realistic. Nevertheless, M-theory, which is the most successful theory unifying in a consistent way gravity and gauge theories, suggests that our world is multi-dimensional. A more elegant and consistent way to realize multi-dimensional spacetimes in quantum field theories is the braneworld scenario \cite{eliplbv,eliplbv1,eliplbv2,eliplbv3,eliplbv4}. Particularly our four dimensional universe can be realized as such a brane, in which matter fields can be localized on the brane or even in some models, propagate in the bulk. Braneworld models can be consistent with observational data when the magnitude of the extra dimensions is quite significant and moreover difficult problems of our four dimensional field theories, such as the cosmological constant and the hierarchy problem, might find explanation within the theoretical framework of brane models. The braneworld corresponds to a five-dimensional manifold with a four dimensional boundary that contains the zero modes of the matter fields. Depending on the model, this five dimensional manifold might be AdS or a more complicated manifold, such the one realized in Randall-Sundrum models. The vacuum energy of the bulk plays a prominent role in such braneworlds, since it contributes to both the brane and bulk cosmological constants. Hence the calculation of the Casimir energy in such models is a compulsory task with very important consequences that their effect affects the spacetime models drastically. Such works have been done in the past and an important stream of papers can be found in \cite{eliplbv,eliplbv1,eliplbv2,eliplbv3,eliplbv4} and references therein.

\noindent With the braneworld scenario being so important, it would be of valuable importance to calculate the Casimir energy, in the case that some dimensions of the total spacetime are of Lifshitz type. Hence we shall discuss in this section, the problem of the scalar field Casimir energy in the case the total spacetime is not flat, and also that there is a conformal coupling between the scalar field and the five dimensional Ricci scalar. Since the total spacetime is not flat, the physics of the problem is contained in the five dimensional Einstein-Hilbert action $\mathcal{S}_H$, plus a scalar matter field part $\mathcal{S}_M$, that is:
\begin{align}\label{matterscalar}
 \mathcal{S}&=\mathcal{S}_H+\mathcal{S}_M\\ \notag & =\int\mathrm{d}^5x\sqrt{-g}\Big{(}M^3R^{(5)}-\Lambda\Big{)}+\frac{1}{2}\int\mathrm{d}^5x\sqrt{-g}\Big{(}-g^{\mu \nu}\partial_{\mu}\phi\partial_{\nu}\phi+\xi R^{(5)}\phi^2\Big{)}
\end{align}
In the above equation, $M$ denotes the five dimensional fundamental mass scale, $\Lambda$ the five dimensional cosmological constant, and $g$ the determinant of the metric. The problem now reduces to that of finding a suitable metric that solves the five dimensional Einstein equations, corresponding to the Einstein-Hilbert action. At this point, let us assume that the total spacetime has Lifshitz extra dimensions, that are scaled as in equation (\ref{scalethree}). Multi-dimensional theories, that assume the scale invariance of equation (\ref{scalethree}), as well as spatial rotation, admit a metric of the form:
\begin{equation}\label{metrscal}
\mathrm{d}s^2=L^2\Big{(}-r^{2z}\mathrm{d}t^2+\frac{\mathrm{d}r^2}{r^2}+r^2\mathrm{d}x^m\mathrm{d}x^m\Big{)}
\end{equation}
where, $L$ is a characteristic length scale. Moreover in the above, $m=1,2,...,d-2$ and $d$ is the total spacetime dimension (which is $d=5$ in our case) and the holographic dimension $r$, scales as $r\rightarrow r/b$. The metric (\ref{metrscal}) is a generalization of the AdS geometry in the case of Lifshitz theories. Thereby, if we assume that the spacetime has Lifshitz scaling in its coordinates, we have to use a metric of the form (\ref{metrscal}). However, the Einstein-Hilbert action (\ref{matterscalar}) cannot admit a Lifshitz geometry of the form (\ref{metrscal}) (see for example \cite{lifn1,lifn2,lifn3,lifn4,lifn5}). Particularly, as shown in \cite{lifn1,lifn2}, the metric (\ref{metrscal}) corresponds to a solution of a modified Einstein-Hilbert action containing a vector and $2-$form, coupled topologically in four dimensions, or two timelike massive vector in $d$-dimensions. This modified action looks like:
\begin{equation}\label{ma}
\mathcal{S}_N=\int\mathrm{d}^{d}x\sqrt{-g}\Big{(}R^{(d)}-\Lambda-\frac{1}{4}F^2+m_A^2A^2\Big{)}
\end{equation}
with $F,A$, massive abelian vector fields. Hence, the task of computing the Casimir energy for a scalar field in Lifshitz invariant spacetimes is not an easy task, in comparison to the flat extra dimensions case. In general, the incorporation of gravity in Lifshitz theories is a difficult but not completely hopeless task. This is because some anti-de Sitter spacetime solutions can be extended to a generalized 4+1 dimensional Horava gravity theory, that may help towards the problem of finding an analogue to Randall-Sundrum metric ansatz. However, this is far beyond the scopes of this article.

\section*{Conclusions}

In this letter we studied the singularities that the Casimir energy of a scalar field has, when Lifshitz dimensions are encountered. Particularly, we were interested in revealing the situations in which the Casimir energy with Lifshitz dimensions has more singularities in reference to the finite Casimir energy of the non-Lifshitz one, and also how the singularities depend on the critical exponents of the Lifshitz dimensions. We addressed the problem in various spacetimes structures.

\noindent Firstly, we studied the massless scalar field Casimir energy in four dimensional spacetime, in which time scales differently in comparison to space dimensions. This case is very well studied in the literature. We confined the scalar field in a box, and having applied Dirichlet boundary conditions at the boundaries of the box, we found that the only problematic situation arises for $z=2$. When $z\geq 3$, the total Casimir energy is finite, a result, that probably stems form the renormalization properties of the Lifshitz quantum field theories. The drawback of the Lifshitz theories is the explicit breaking of Lorentz invariance, but in
the case where boundaries are introduced, there is no room for Lorentz invariance anyway. Additionally, we studied the massless scalar field confined in a box, but this time with the space dimensions having different critical exponents and, the Lifshitz dimensions are the three space dimensions. In this case, the results are similar as in the previous case, that is, when $z_3\geq 1$ and $z_1=z_2=2$, the Casimir energy is finite.

\noindent In addition, we studied the massless scalar field in Minkowski spacetime with one and two circular extra dimensions, with the scalar field obeying periodic boundary conditions in the extra dimensions. In the case of one extra dimension we separated the 4-dimensional part of the Casimir energy from the extra dimensional contribution to the energy. We found that when the critical exponent is $z_1\geq 2$, the Casimir energy is singular, a feature that the non-Lifshitz Casimir energy does not have. In the case of two extra dimensions, we found similar results. Specifically, when the critical exponents of each dimension separately satisfy $z_i\geq 2$, the massless scalar field Casimir energy is singular.

\noindent Lastly, we briefly addressed the calculation of the Casimir energy within the context of braneworld scenario, in the case that Lifshitz dimensions exist in the curved spacetime. As we demonstrated, such a calculation is not an easy task, since the Lifshitz metric is not a solution of the Einstein equations. Moreover, the addition of massive scalar fields is required in order the Lifshitz metric satisfies the Einstein equations. Hence, the study of Casimir energy gets even more complicated.

\noindent Since the Casimir energy is an important feature of a consistent quantum field theory, being a manifestation of the vacuum of the theory, the study of the singularities that this energy exhibits is of particular importance. This is owing to the fact that any infinite Casimir energy term must be properly regularized in a physically consistent way. Therefore, within the theoretical framework of Lifshitz spacetime dimensions with the total number of dimensions being equal to four, there are values of the critical exponents for which, the Casimir energy is finite, as in the non-Lifshitz case. Moreover, the only value of the critical exponent, that yields a singular Casimir energy, even in the context of zeta-regularization, is $z=2$. In the case that Lifshitz extra dimensions are considered, the Casimir energy contains singularities, in contrast to the non-Lifshitz case.

\noindent In conclusion, 4-dimensional quantum field theories with Lifshitz dimensions and $z\geq 3$, are physically appealing, while theories with Lifshitz extra dimensions case are less appealing, compared to the non-Lifshitz case owing to the fact that we have to give sufficient explanations regarding the origin of the new singularities, and of course we have to properly eliminate these.

\noindent It would be very interesting to extend the above analysis in the finite temperature case, or studying the fermionic Casimir energy at zero and finite temperature, with or without extra dimensions. Particularly, a natural question to ask is how the bag boundary conditions affect the Casimir energy within the theoretical framework of Lifshitz dimensions. Of equal importance is the electromagnetic field study of the Casimir energy with Lifshitz dimensions, and how the use of different gauges modifies the final result. We hope to address these problems in the future.

\end{document}